# scPyviewer: a Python-native interactive viewer from AnnData single-cell data

Hao Xuan[1,*], Yu Huang[1,2], Jiang Bian[1,2,3], Xiangtao Liu[4]

[1]Department of Biostatistics and Health Data Science, Indiana University School of Medicine, Indianapolis, IN 46202, USA

[2]Center for Biomedical Informatics, Regenstrief Institute, Indianapolis, IN 46202, USA

[3]Indiana University Health, Indianapolis, IN 46290, USA

[4]Department of Psychiatry, University of Iowa, Iowa City, IA 52242, USA

*Correspondence: xuanhao@iu.edu (H.X.)

## Abstract

**Motivation:** Interactive tools that let non-programmers explore an analyzed single-cell dataset, its embeddings, gene expression, cell metadata, and marker genes, have become standard laboratory infrastructure. Every actively maintained tool in this space (ShinyCell, ScRDAVis, sCIRCLE, scViewer) is built on R Shiny and requires a Seurat object as input. Laboratories whose primary analysis pipeline is Python/scanpy, the dominant framework for single-cell RNA-seq, spatial, and multi-omic analysis, therefore have no lightweight, language-native option that pairs a shareable web-based viewer with a scriptable Python API: sharing a scanpy result means either exporting to Seurat first or handing over a notebook that only a programmer can run.

**Results:** We present scPyviewer, a web-based viewer that ingests AnnData objects directly and reproduces the core interaction patterns of the incumbent R Shiny tools without leaving the Python stack. In a feature-parity audit against three actively maintained R Shiny incumbents, scPyviewer matches or exceeds every baseline capability (7/7); among these, it uniquely offers native AnnData ingestion with no Seurat conversion, and cross-dataset comparison over shared genes and matched cell-type composition. Benchmarked head-to-head against the R/Seurat rendering substrate the incumbents are built on, identical operations, identical data, across three datasets spanning 22,315 to roughly 313,000 cells, scPyviewer renders every core view faster at every scale tested (up to 3.6× on a single view) and at a fraction of the memory (5.2× lower on the smallest dataset). At the largest scale tested, the gap becomes categorical rather than incremental: scPyviewer completes every view on a 313,000-cell dataset while the Seurat substrate exhausts an 8 GB memory budget and fails outright. Beyond the interactive app, scPyviewer installs via pip or conda and exposes a public Python API that returns Matplotlib figures and pandas tables for scripted, publication-ready output.

**Availability and implementation**: scPyviewer is implemented in Python 3.11 (scanpy 1.11.5, anndata 0.12.19, streamlit 1.59.2, plotly 6.9.0) and distributed with a one-command reproduction interface that installs pinned dependencies, regenerates the benchmark and all figures, and launches the interactive app. Source code is available at https://github.com/xuan13hao/scPyviewer.git.

## Background

Single-cell genomics routinely produces datasets that bench scientists need to interrogate long after computational analysis is finished, which cells express a gene of interest, how does its expression vary across conditions, and what defines a given cluster? Answering these questions interactively, without writing code, is the purpose of "cell browser" tools that load pre-analyzed datasets and expose embeddings, expression overlays, and marker tables through a web user interface. The established maintained tools in this category, ShinyCell [1], ScRDAVis [2], sCIRCLE [3] and scViewer [4], share an architectural commitment: they are R Shiny applications built on Seurat objects [5]. The UCSC Cell Browser similarly serves this need, enabling wet-lab scientists who generated the samples to explore embeddings, marker genes, and cell-type calls independently of the computational analyst [6]. Consensus best-practice guidance explicitly recognizes this handoff, letting collaborators query a finished, analyzed dataset without writing code, as a formal downstream step of single-cell analysis, not an afterthought [7].

Yet this R-centric ecosystem is becoming a poor fit for a large and growing fraction of the field. Seurat popularized the single in-memory object that bundles counts, normalized expression, embeddings, and metadata, and the Shiny ecosystem grew directly from that convenience. Tracked across the field's history, however, the trend among newly released single-cell software runs in the opposite direction: Python's share has risen steadily while R's has fallen, with crossover projected to establish Python as the dominant implementation language for single-cell analysis software [8]. At the center of this shift sit scanpy [6] and its AnnData container [7], which have become the shared substrate not only for core preprocessing and clustering but for spatial-omics toolkits, deep generative models, and foundation-model cell embeddings that increasingly extend analyses beyond a single assay. For Python-first groups, using an R Shiny viewer imposes an explicit, error-prone AnnData-to-Seurat conversion, and every newer, Python-only output is invisible to a Seurat-based viewer unless it is exported from the ecosystem that generated it. Despite the maturity of Python web frameworks such as Streamlit [9] and Dash, no actively maintained, shareable, no-code browser reads AnnData natively without a Seurat dependency; two related but architecturally narrower Python-facing efforts, SCSEQ [10] and Sciviewer [11], are situated relative to scPyviewer in the Discussion.

Building on this gap, our aim is a Python-native cell browser that reproduces the interaction patterns bench scientists already expect from the R Shiny incumbents while reading AnnData objects natively, with no conversion step, and remaining usable as datasets outgrow what a single desktop can hold in memory. We address this with scPyviewer, a Streamlit application organized around a framework-independent data-access layer and plotting layer (Methods), so that the same tested code serves the interactive app, the benchmark suite, and a scripted Python API alike. Concretely, this paper makes four contributions: (1) a Python/AnnData-native cell browser that matches every core capability of the maintained R Shiny incumbents (ShinyCell, ScRDAVis, sCIRCLE/scViewer) while adding native AnnData ingestion and cross-dataset/cross-species comparison, neither offered by any of them; (2) a guarded, idempotent preprocessing pipeline that makes an arbitrary analyzed AnnData object viewer-ready without overwriting existing work; (3) a like-for-like, cross-language performance evaluation against the shared Seurat rendering substrate the incumbents depend on, across three species and nearly a fifteen-fold range of dataset size, showing scPyviewer faster at every scale tested and the only stack that completes every view at 313,000 cells on an 8 GB host; and (4) a public Python API that exposes the same plotting and data-access functions as Matplotlib figures and pandas tables for reproducible, scripted output outside the interactive app.

## Methods

### Architecture

scPyviewer is organized as a small, layered Python package with no framework code in its plotting or data layers, which keeps those layers unit-testable and reusable. A dedicated data-access layer handles dataset discovery and reads directly from AnnData objects, exposing operations to load a dataset, retrieve cell-level metadata, extract two-dimensional embeddings, pull per-gene expression vectors, assign consistent colors to categorical variables, retrieve the marker and differential-expression table, and search for genes by name, all without any dependency on the Streamlit framework. A separate plotting layer provides a set of pure Plotly [12] plotting functions covering embedding scatter plots, multi-gene expression grids, violin plots, dot plots, and composition bar charts; each function takes simple arrays as input and returns a figure object, with no reliance on global state. These two layers are consumed by the interactive Streamlit application itself, which presents a sidebar dataset picker and shared metadata filters alongside five tabs, Embedding, Expression, Markers/DE, Compare, and Export, and loads the analyzed object once, caching it in memory for the duration of the session.

### AnnData ingestion and preprocessing

scPyviewer reads AnnData files directly; there is no conversion step. To make an arbitrary analyzed (or partially analyzed) object ready, a preprocessing command-line tool runs a guarded, idempotent pipeline: every step checks whether its output already exists and skips it if so, so re-running the tool only fills in what is missing and never overwrites an existing analysis. The pipeline first preserves the raw count matrix, builds a log-normalized expression matrix, and sets it as the primary matrix used for downstream analysis, skipping this step entirely if the data already appear to be log-normalized, as judged by a heuristic on the range and integer-ness of the values. It then performs highly-variable-gene selection followed by principal component analysis, skipping this step if a principal- component embedding is already present. Next, it computes UMAP [13] and t-SNE [14] embeddings while preserving any embedding that is already present, on the demonstration dataset, a precomputed UMAP embedding and a 1,280-dimension foundation-model embedding from the Universal Cell Embedding model [15] were both retained, and UMAP computation was correctly skipped. The pipeline then performs per-group differential expression using a Wilcoxon rank-sum test [16] over an automatically selected grouping column, a categorical metadata field such as cell type, storing the full results and flattening them into a tidy marker table for the browser. Finally, it records a provenance block summarizing what the pipeline did, alongside a sidecar manifest file describing the prepared object. The pipeline is deliberately species- and assay-agnostic: it auto-selects the grouping column and never hard-codes gene names or organism assumptions.

**Feature modules**

The four analytical modules map directly onto the interaction patterns the incumbent tools established, plus a comparison module they lack. The Embedding module displays any two-dimensional embedding colored either by a metadata column or by the expression of a chosen gene. The Expression module provides single- and multi-gene overlays on the embedding, together with violin and dot plots grouped by any metadata column. The Markers/DE module allows interactive browsing of the per-group differential-expression table. The Compare module offers side-by-side views across datasets or species over shared genes and matched cell-type composition, a capability absent from all three incumbents. Finally, the Export module allows the current view and the filtered cell table to be downloaded.

**Benchmark design**

We evaluated scPyviewer along four axes, all reproducible through a single benchmarking command. The first, feature parity, is a checklist of seven capabilities drawn from the published feature sets of ShinyCell, ScRDAVis, and sCIRCLE/scViewer, scored for each tool and for scPyviewer, with a code reference recorded as evidence for every scPyviewer entry. The second, performance, comprises wall-clock load time and render latency for each core view (best and mean of three runs), plus peak Python memory measured via built-in memory-profiling instrumentation, all recorded on the demonstration dataset. The third, a cross-language head-to-head, times the identical operations on the identical data against the R/Seurat rendering substrate that the incumbent tools are built on: ShinyCell, ScRDAVis, sCIRCLE, and scViewer all render through the same underlying Seurat plotting functions for embeddings, feature overlays, violin plots, and dot plots, plus a ggplot2-based composition bar, so we benchmark that shared substrate directly (best-of-three render, single-run load, peak resident-set-size memory via psutil) rather than launching a full Shiny server, which would fold in framework-specific web overhead orthogonal to the rendering cost itself; this isolates a like-for-like comparison of the work each stack performs to draw a view. Because scPyviewer's memory is captured with an in-process Python profiler and the Seurat substrate's memory is captured as OS-level resident-set-size (RSS) via psutil, the two measurements are not strictly equivalent instruments; each was, however, applied consistently within its own stack across all three datasets, and the reported memory gap should be read as a large, consistent difference rather than a precisely calibrated ratio. We ran this head-to-head across three datasets spanning roughly an order of magnitude in size (22,315 to approximately 313,000 cells; see Datasets, below) to test not only relative speed but whether each stack remains usable as scale increases, on a fixed 16-CPU, 8 GiB, no-GPU host. A run was scored as an out-of-memory (OOM) failure if the process was killed or failed to complete a view within the 8 GB host memory budget; scPyviewer completed every run without triggering this condition, so OOM was observed only for the Seurat substrate.

**Datasets**

We selected three publicly available datasets that together span three species, three tissues, two experimental designs (developmental trajectory and disease comparison), and an order-of-magnitude range in dataset size, from 22K to 313K cells, to test the claim that scPyviewer’s pipeline and viewer make no organism- or assay-specific assumptions (Table 1).

For chicken-heart developmental atlas (Gallus gallus, 22,315 cells × 10,031 genes) [17], single-cell RNA-seq profiling of the developing chicken heart across seven samples spanning four time points: embryonic day 4 (D4), D7, D10, and D14, with left-ventricle/right-ventricle split at D7–D14. Fifteen annotated cell types cover the major cardiac lineages (two cardiomyocyte populations, fibroblasts, endocardial and vascular endothelial cells, mural cells, valve cells, erythrocytes, macrophages, dendritic cells, and several progenitor/transitional populations). This dataset serves as the primary benchmark for single-dataset performance and the cross-language head-to-head; it represents a typical mid-sized atlas that fits in memory and exercises the full feature set.
On green monkey lung and mediastinal lymph node (Chlorocebus sabaeus, 78,440 cells × 13,389 genes), single-cell RNA-seq profiling of two anatomically paired tissues, lung parenchyma and mediastinal lymph node, from ten subjects [18]. Eighteen annotated cell types include major immune populations (T cells, B cells, NK cells, macrophages, monocytes, dendritic cells, plasma cells, neutrophils, mast cells, ILCs, pDCs), structural lung populations (pneumocytes, ciliated cells, fibroblasts, vessel-associated cells), and a dividing/progenitor cluster.
A large-scale single-cell survey of human lung disease comparing three conditions: idiopathic pulmonary fibrosis (IPF), chronic obstructive pulmonary disease (COPD), and healthy controls [19]. Thirty-nine fine-grained cell identities are assigned to six broad compartments: Epithelial, Endothelial, Myeloid, Lymphoid, Stromal, and Multiplet. At 6 GB on disk, the .h5ad file exceeds the in-memory threshold and is opened in backed mode; this makes it the stress test for both the preprocessing pipeline on large files and the viewer's metadata/embedding operations at 14× the primary dataset scale.

## Results

### Feature parity: scPyviewer matches every baseline capability and adds two

Across the seven-item capability checklist, scPyviewer scores “Yes” on every item in Figure 1. All four tools support basic embedding plots colored by metadata and no-code shareable deployment. ShinyCell and ScRDAVis fully support multi-gene expression overlays and violin/box plots grouped by metadata, while sCIRCLE/scViewer supports both only partially; scPyviewer matches the strongest incumbents on both. Marker/DE table browsing is fully supported by ScRDAVis and by scPyviewer, only partially by ShinyCell, and not at all by sCIRCLE/scViewer. scPyviewer is also the only tool to offer two further capabilities that no incumbent provides in any form: native Python/AnnData input that requires no Seurat conversion, and cross-dataset/cross-species comparison.

### Performance: all core views

We benchmarked scPyviewer against the R/Seurat rendering substrate the incumbents are built on, identical operations, identical data, across three datasets of increasing size: the 22,315-cell chicken-heart developmental atlas, a 78,000-cell green-monkey dataset, and a 313,000-cell human-lung dataset, all measured on the same 16-CPU, 8 GiB, no-GPU host.

On the chicken-heart dataset (Figure 2A), scPyviewer loads the prepared object in 0.63 s against 3.52 s for Seurat (5.6×) and renders every core view faster except the composition bar, where Seurat has a slight edge (0.17 s vs. 0.22 s): gene-colored embedding 0.12 s vs. 0.44 s (3.6×), metadata-colored embedding 0.24 s vs. 0.50 s (2.1×), multi-gene grid 0.41 s vs. 1.08 s (2.6×), violin 0.26 s vs. 0.48 s (1.9×), and dot plot 0.14 s vs. 0.34 s (2.4×). The same pattern holds on the larger, 78,000-cell green-monkey dataset (Figure 2B): load time 0.95 s vs. 2.94 s (3.1×), and every core view 1.5–3.4× faster, with the composition bar again close to parity (0.9×).

At the largest scale tested, a 313,000-cell human-lung dataset, the comparison stops being a matter of degree. scPyviewer loads the dataset and renders every view (load 7.43 s; gene-colored embedding 0.24 s; metadata-colored embedding 0.52 s; multi-gene grid 1.92 s; violin 0.67 s; dot plot 0.40 s; composition bar 0.41 s), while the Seurat substrate runs out of memory and fails to complete a single view on the same 8 GB host (Figure 2C). Peak resident memory across all three datasets (Figure 2D) shows why: scPyviewer's memory footprint grows roughly in step with dataset size (206 MB at 22K cells, 637 MB at 78K cells, 6,350 MB at 313K cells), while Seurat's footprint grows faster (1,076 MB, 5.2× scPyviewer's, at 22K cells; 2,019 MB, 3.2×, at 78K cells) and exceeds the 8 GB machine limit before the human-lung dataset can be processed at all. scPyviewer is therefore not only faster at every scale we tested but remains usable at a scale where the Seurat substrate the incumbent tools depend on is not.

**Demonstration: the live tool on a developmental atlas**

The annotated UMAP resolves fine-grained cell states across the atlas (Figure 3): a cardiomyocyte compartment split into immature, MT-enriched, and a second cardiomyocyte subpopulation; a connected mesenchymal arm covering epicardial-mesenchymal, epicardial-epithelial, mural, fibroblast, valve, endocardial, and vascular endothelial cells; and two well-separated immune/blood islands, macrophages and dendritic cells on one side, erythrocytes on the other. The same Embedding view resolves an equally coherent cell-type landscape on the human-lung dataset, shown at both fine-grained and broad-compartment resolution (Supplementary Figure.S1, S2), confirming that cluster structure remains interpretable at an order of magnitude more cells and across species. Grouped by sample, a representative marker gene's expression varies by developmental stage and cardiac chamber (Figure 4): expression rises from day 4 (D4, before chamber septation is complete) through day 7 (D7_LV, D7_RV) and falls again by day 14 (D14_LV, D14_RV), with left- and right-ventricle samples tracking closely at each stage, the kind of sample- and metadata-grouped view the Expression module is built to surface. The same violin view is reproduced on the green-monkey and human-lung datasets (Supplementary Figure. S3, S4), and the chicken-heart dataset additionally anchors a sample-level composition view not shown in the main text (Supplementary Figure. S5).

Multi-gene expression overlays place canonical markers in the expected compartments (Figure 5): the cardiac myosin MYL2 is concentrated in the cardiomyocyte compartment, the hemoglobin HBA1 is restricted almost entirely to the separate erythrocyte island, and the matrix genes FN1, MDK, and POSTN mark the mesenchymal arm, consistent with the fibroblast, mural, and valve populations annotated in Figure 3. IFI6 shows negligible expression across the atlas, illustrating that the same multi-gene grid also makes an absence of signal immediately legible, not only a hit. The Markers/DE module's underlying table view, not shown in the main text for the chicken-heart dataset, is illustrated there and reproduced on the green-monkey and human-lung datasets alike (Supplementary Figure. S6-S8), completing the demonstration across all three species.

## Discussion

scPyviewer demonstrates that the core interaction patterns of the mature R Shiny cell-browser ecosystem can be reproduced natively in Python, removing the language switch and the AnnData-to-Seurat conversion that Python-first labs currently pay. It matches the incumbents on every baseline capability and adds native AnnData ingestion and cross-dataset/cross-species comparison, at render latencies that keep interaction immediate across three species and nearly a fifteen-fold range of dataset sizes. The most consequential result is not the per-view speedup but what happens at the top of that range: on the 313,000-cell human-lung dataset, the Seurat substrate the incumbent tools depend on runs out of memory and fails outright, while scPyviewer completes every view on the same 8 GB host. For labs whose datasets are growing faster than their hardware budgets, that is a difference in kind, not just degree, a dataset that an R Shiny incumbent cannot open at all is one scPyviewer renders in under two seconds per view.

The quantitative results now span three datasets, three species, and roughly an order of magnitude in cell count. The three datasets used here make the Compare module's cross-species capability testable for the first time, but we have not yet reported a dedicated stress test of that module on genuinely disparate inputs (differing gene sets, annotation vocabularies, and normalization conventions); that evaluation, rather than raw data availability, is what remains future work. The performance figures are also machine-specific (measured on a 16-CPU, 8 GiB, no-GPU host) and will shift on other hardware, though the OOM result is a property of the 8 GB budget relative to Seurat's memory growth rather than an artifact of this specific machine, and we would expect it to recur on any similarly provisioned host. The cross-language comparison benchmarks the shared Seurat rendering substrate the incumbents build on rather than each Shiny app end-to-end; a full app-to-app comparison would fold in framework-specific web overhead but could not make the R tools render faster, or survive further past the memory ceiling, than the Seurat functions they call.

We built the reference implementation in Streamlit for its minimal boilerplate and one-command deployment, which suits a lab tool meant to be launched and shared quickly. Dash offers finer control over callback graphs and client-side state, which would matter for very large datasets or more complex linked-view interactions; because scPyviewer's plotting and data layers are framework-independent pure functions, a Dash front end could reuse them directly. Render latencies stayed well within interactive range across the full tested scale, from 22K to 313K cells, so the added complexity of Dash was not warranted here, though as datasets grow further, the same headroom argument will eventually need re-testing rather than assumed.

Two narrower Python-facing efforts are worth situating scPyviewer against, based on their published descriptions rather than a formal benchmark performed here. SCSEQ provides a Python web front end but computes on Seurat objects server-side and requires uploading data to a hosted service, so it retains the Seurat dependency and external-data-sharing requirement that scPyviewer is designed to avoid. Sciviewer operates natively on Python/AnnData objects but is launched as an embedding-inspection window from within an analyst's own Jupyter session; it is not packaged as a standalone application a non-programming collaborator can open independently, which is the specific capability scPyviewer's Streamlit deployment targets. These are architectural distinctions verifiable from each tool's own description rather than claims requiring a head-to-head benchmark; a formal feature-parity or performance comparison against both tools is a natural extension of the present work.

The human-lung dataset already forced part of this question in practice rather than in principle: at roughly 6 GB on disk, it exceeded the in-memory threshold we set and was instead opened in AnnData's backed mode, and the pipeline and viewer handled it without code changes. That is direct evidence, not just a plausible extrapolation, that the on-disk/backed-access strategy we previously described as future work is viable for at least this scale and file size. What remains open is server-side downsampling for the scatter views at even larger scale, and characterizing where backed-mode access itself starts to cost interactivity as datasets grow past what we tested here; the pure-function plotting layer is compatible with such optimizations without changes to the UI.

## Figure legends and table

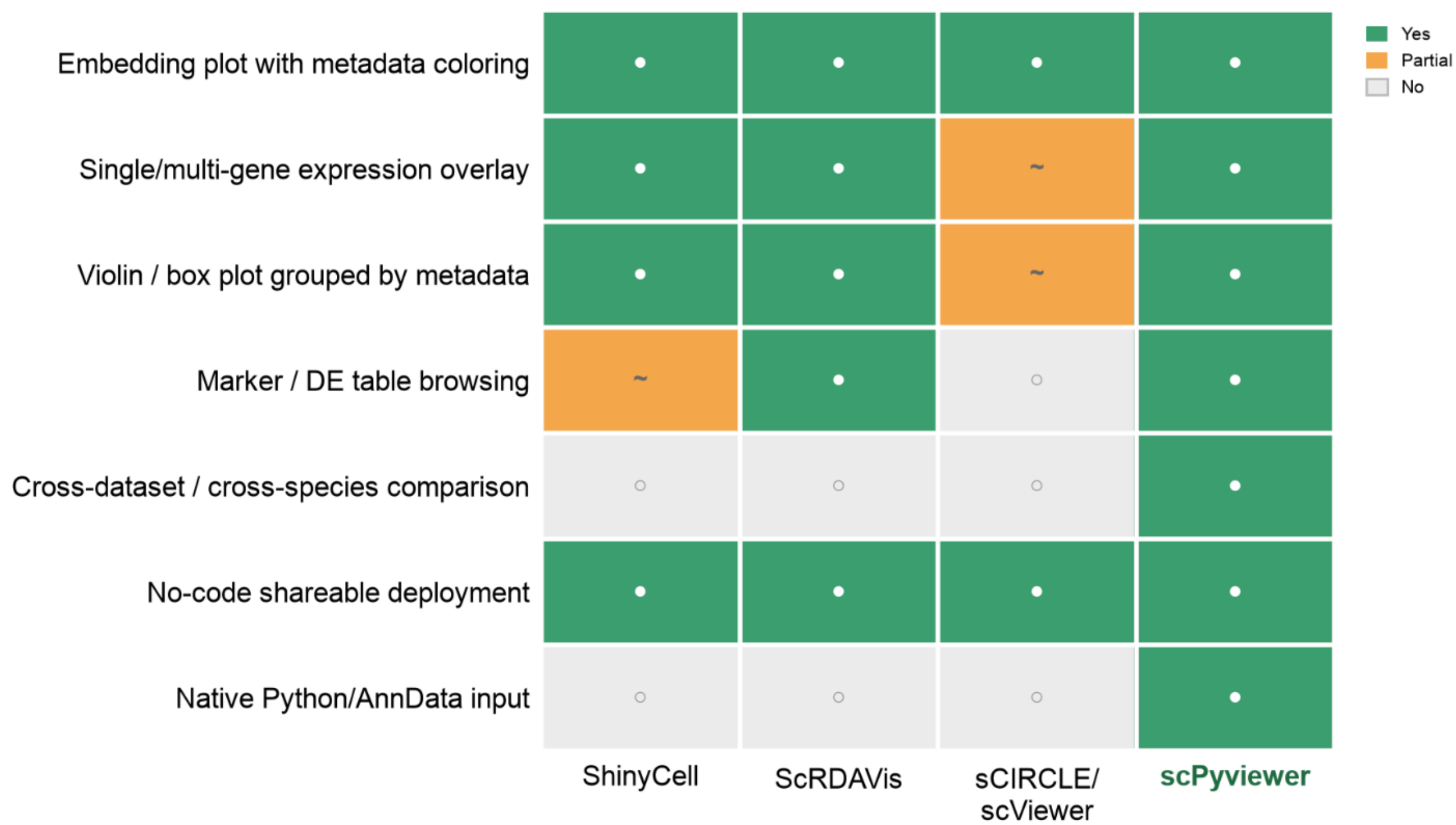

**Figure 1:** Feature-parity matrix. Support level (green, Yes; orange, Partial; grey, No) for seven core capabilities across three R Shiny incumbents (ShinyCell, ScRDAVis, sCIRCLE/scViewer) and scPyviewer. scPyviewer is the only tool scoring Yes on all seven, including native AnnData ingestion and cross-dataset/cross-species comparison.

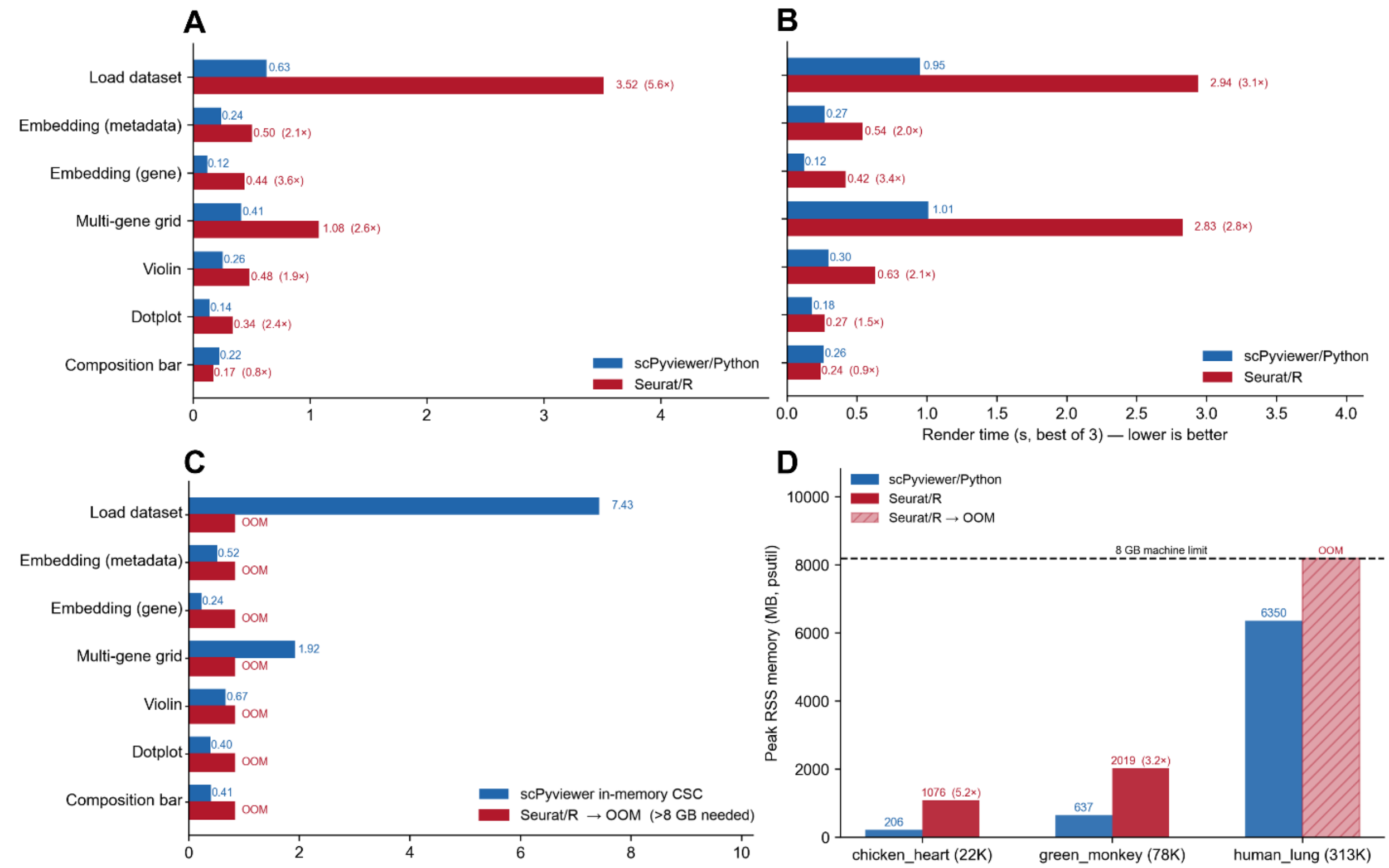

**Figure 2:** scPyviewer outpaces the Seurat rendering substrate across three orders of magnitude in dataset size. **a**, Render time per view, chicken-heart dataset (22,315 cells), scPyviewer vs. Seurat. **b**, As in a, green-monkey dataset (78,440 cells). **c**, Render time for scPyviewer on the human-lung dataset (312,928 cells); Seurat fails with an out-of-memory (OOM) error on every view and is not plotted. **d**, Peak RSS memory across all three datasets; dashed line, 8 GB host limit, exceeded by Seurat (OOM) at 312,928 cells. Bars show best of three replicates on the same 16-CPU, 8 GB, no-GPU host.

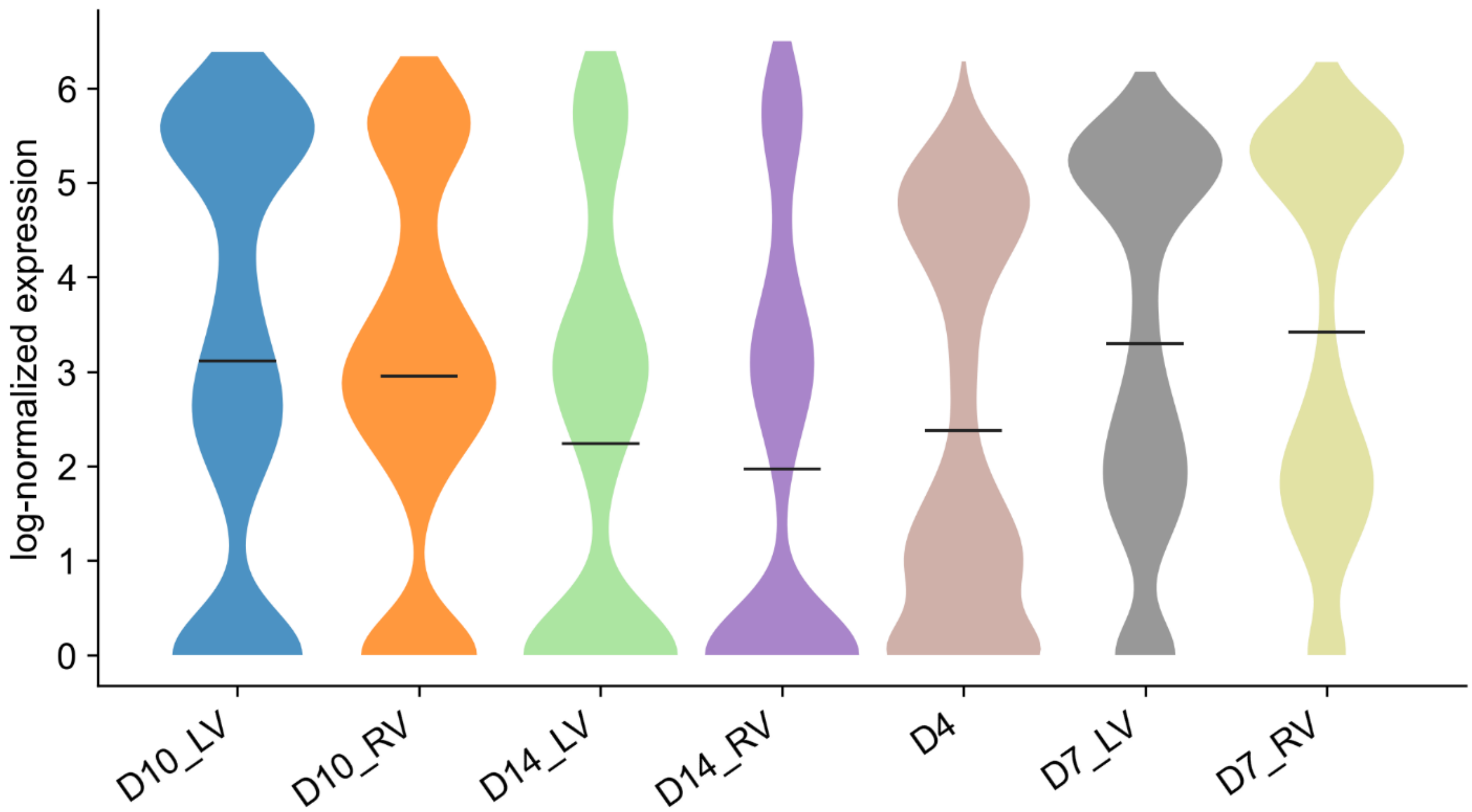

**Figure 3:** Cell-type landscape of the developing chicken heart. UMAP of 22,315 cells colored by annotated cell type, resolving cardiomyocyte, mesenchymal, and immune/blood lineages.

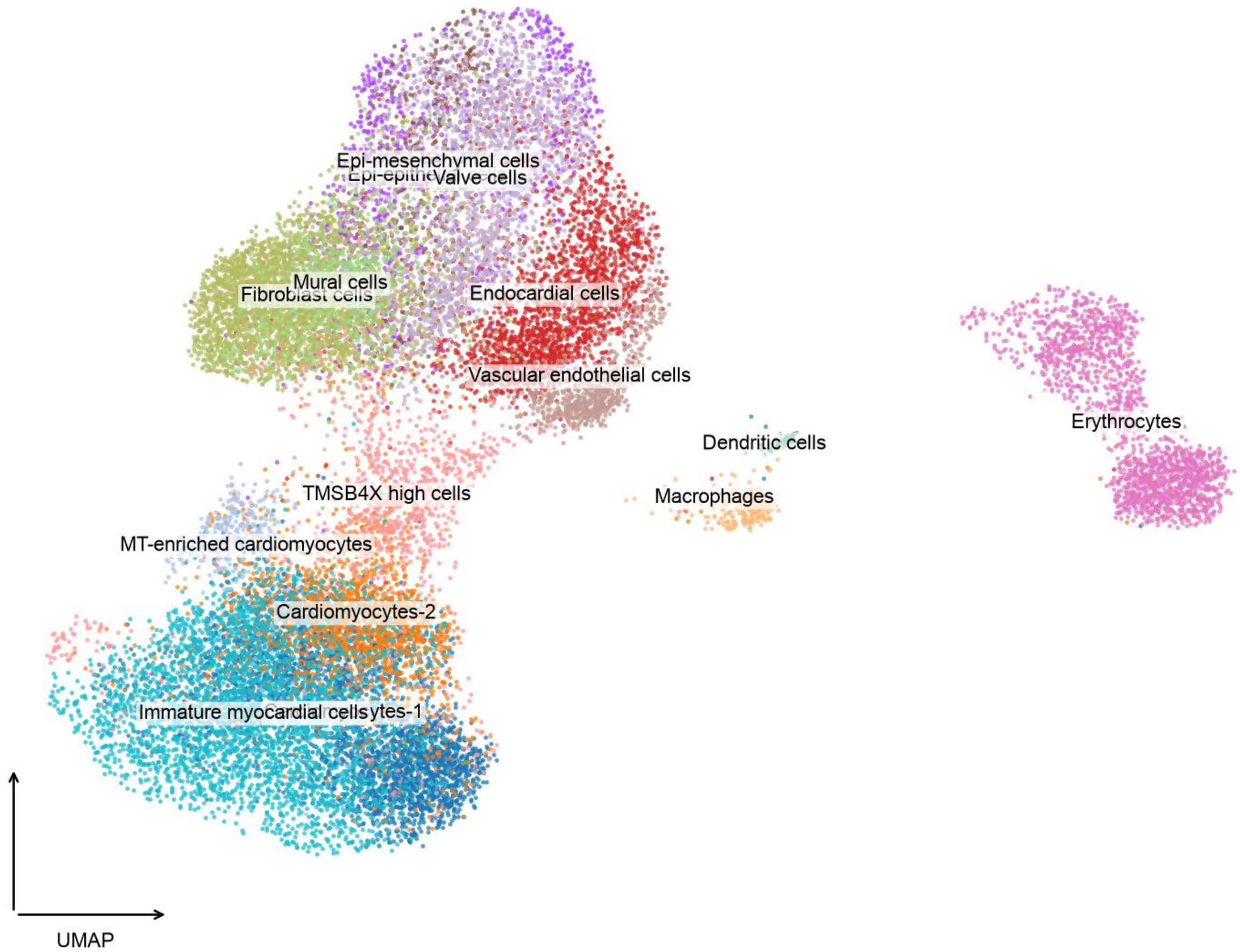

**Figure 4:** Sample-level expression. Violin plots of log-normalized expression across seven chicken-heart samples (D4; D7, D10, D14 each split into LV/RV). Bars, sample means.

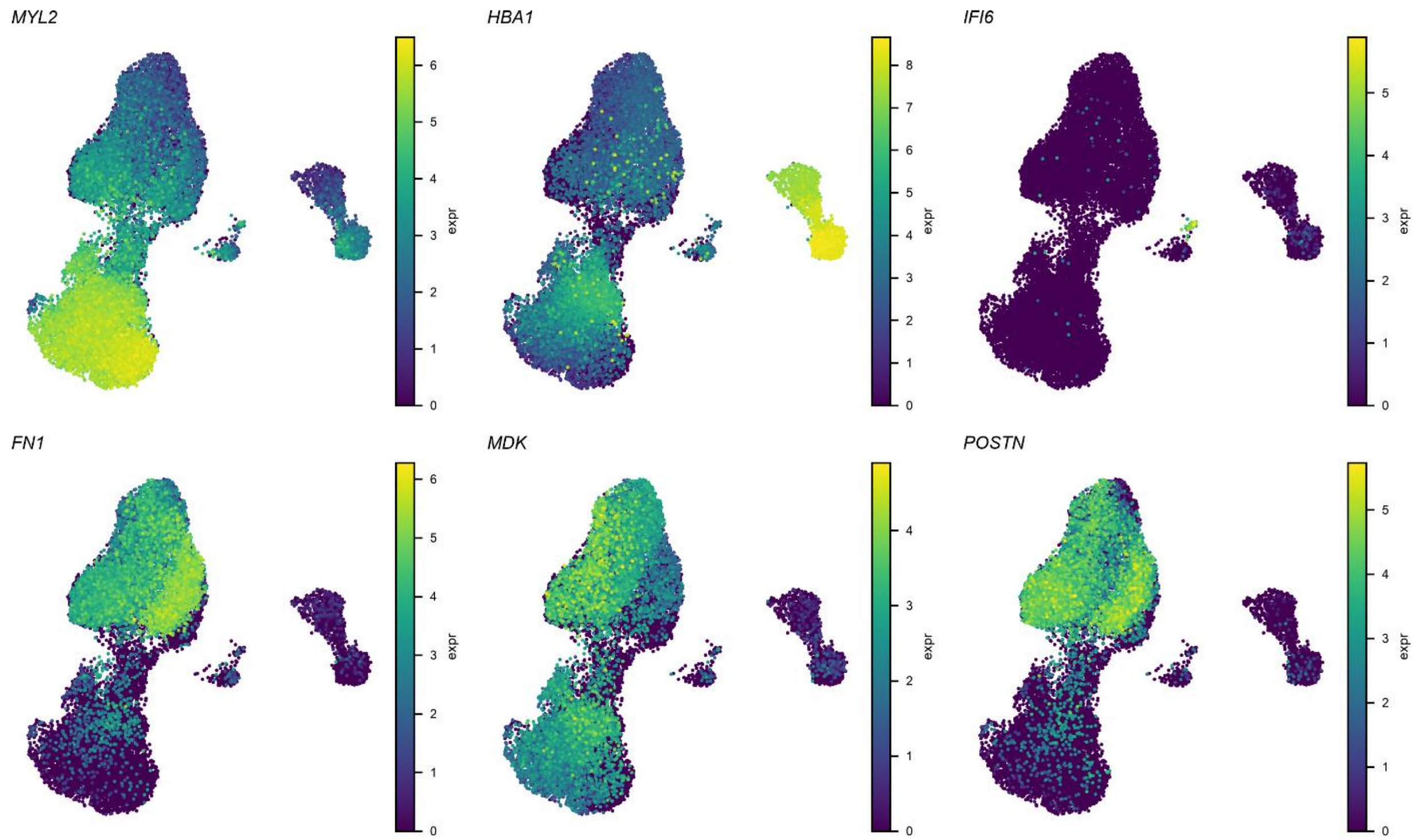

**Figure 5:** Lineage-specific marker expression. UMAP overlays colored by log-normalized expressions of MYL2, HBA1, IFI6, FN1, MDK, and POSTN. MYL2, cardiomyocytes; HBA1, erythrocytes; FN1/MDK/POSTN, mesenchyme; IFI6, no detectable signal.

**Table 1**: Summary of the three datasets used in this study.

| Dataset | Species | Cells | Genes | Cell types | Groups |
|---|---|---|---|---|---|
| Chicken heart | *Gallus gallus* | 22,315 | 10,031 | 15 | 7 samples (D4–D14) |
| Green monkey lung/LN | *Chlorocebus sabaeus* | 78,440 | 13,389 | 18 | 10 subjects × 2 tissues |
| Human lung disease | *Homo sapiens* | 312,928 | 18,089 | 39 (6 broad) | 3 disease conditions |

## Funding

This research received no specific grant from any funding agency in the public, commercial, or not-for-profit sectors.

## Competing Interests

All other authors declare that they have no competing interests.

## Author Contributions

Conceptualization was carried out by HX; methodology was developed by HX. HX developed the software and generated the visualizations. Formal analysis was performed by HX; investigation was conducted by HX. Project supervision was provided by HX and YH. The original draft was written by HX, and the final manuscript was reviewed and edited by HX, YH, XL, and JB.

**Supplementary Materials for**

**“scPyviewer: a Python-native interactive viewer from AnnData single-cell data”**

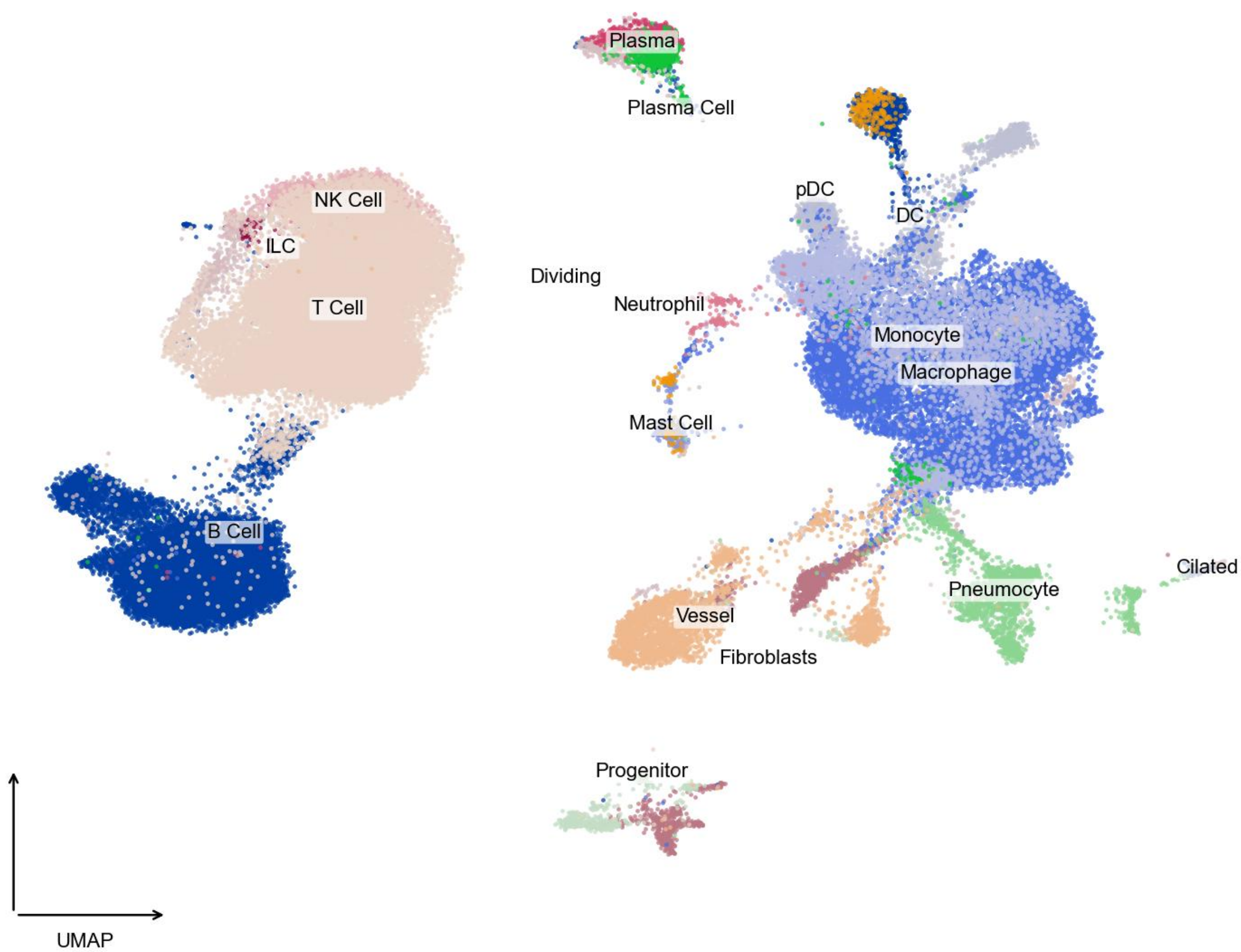

Figure S6: Cell-type landscape of the human-lung dataset. UMAP of 312,928 cells colored by annotated cell type, resolving lymphoid (T cell, B cell, NK cell, ILC), myeloid (monocyte, macrophage, dendritic cell, pDC, neutrophil, mast cell), plasma/plasma cell, epithelial (pneumocyte, ciliated), stromal (fibroblast, vessel), and progenitor/dividing populations. Generated from the same 312,928-cell dataset used in the scalability benchmark.

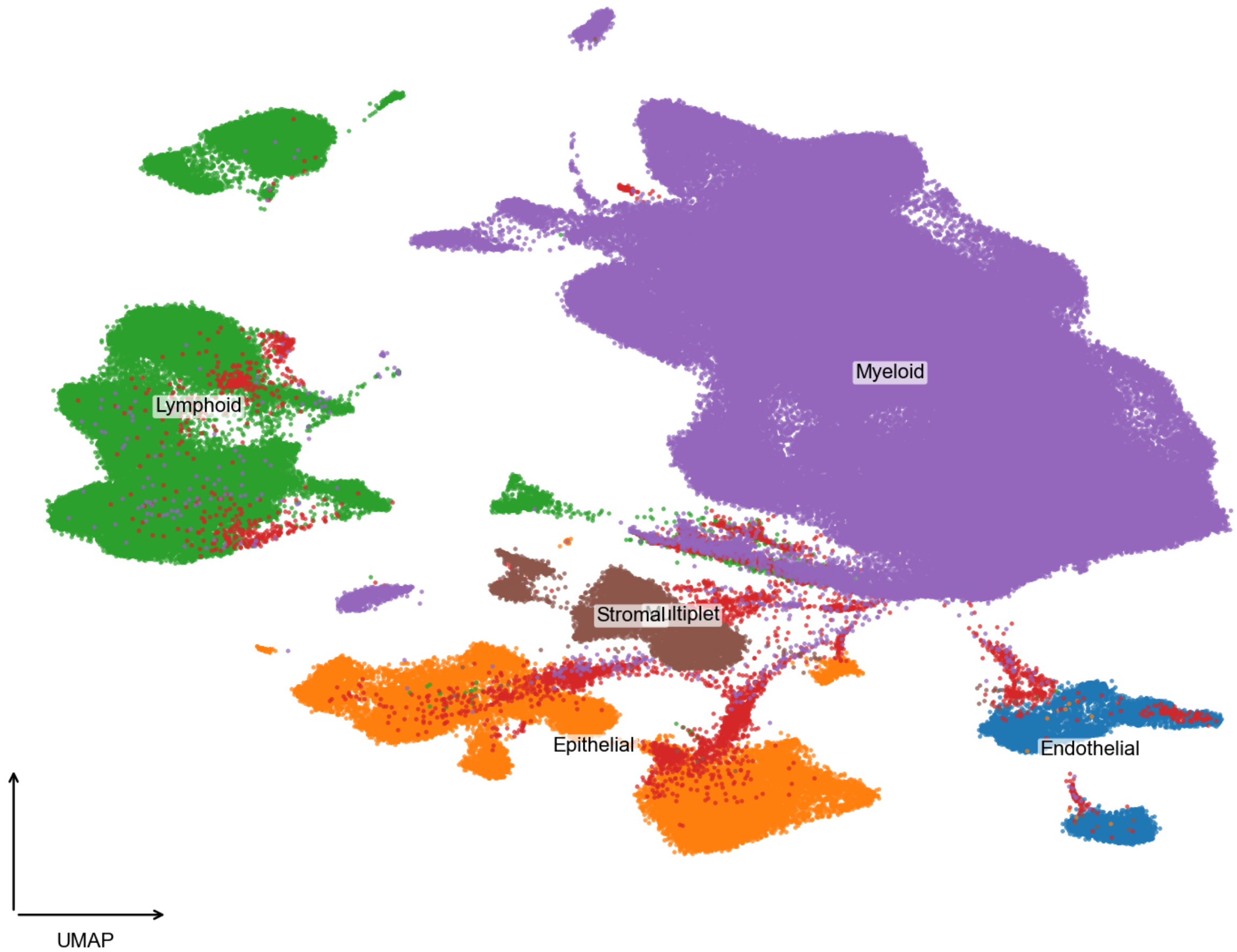

Figure S7: Human-lung cell-type atlas at two resolutions. UMAP of 312,928 cells colored by fine-grained cell identity (39 populations, previous figure) and by the six broad compartments, Myeloid, Lymphoid, Epithelial, Endothelial, Stromal, and Multiplet.

d

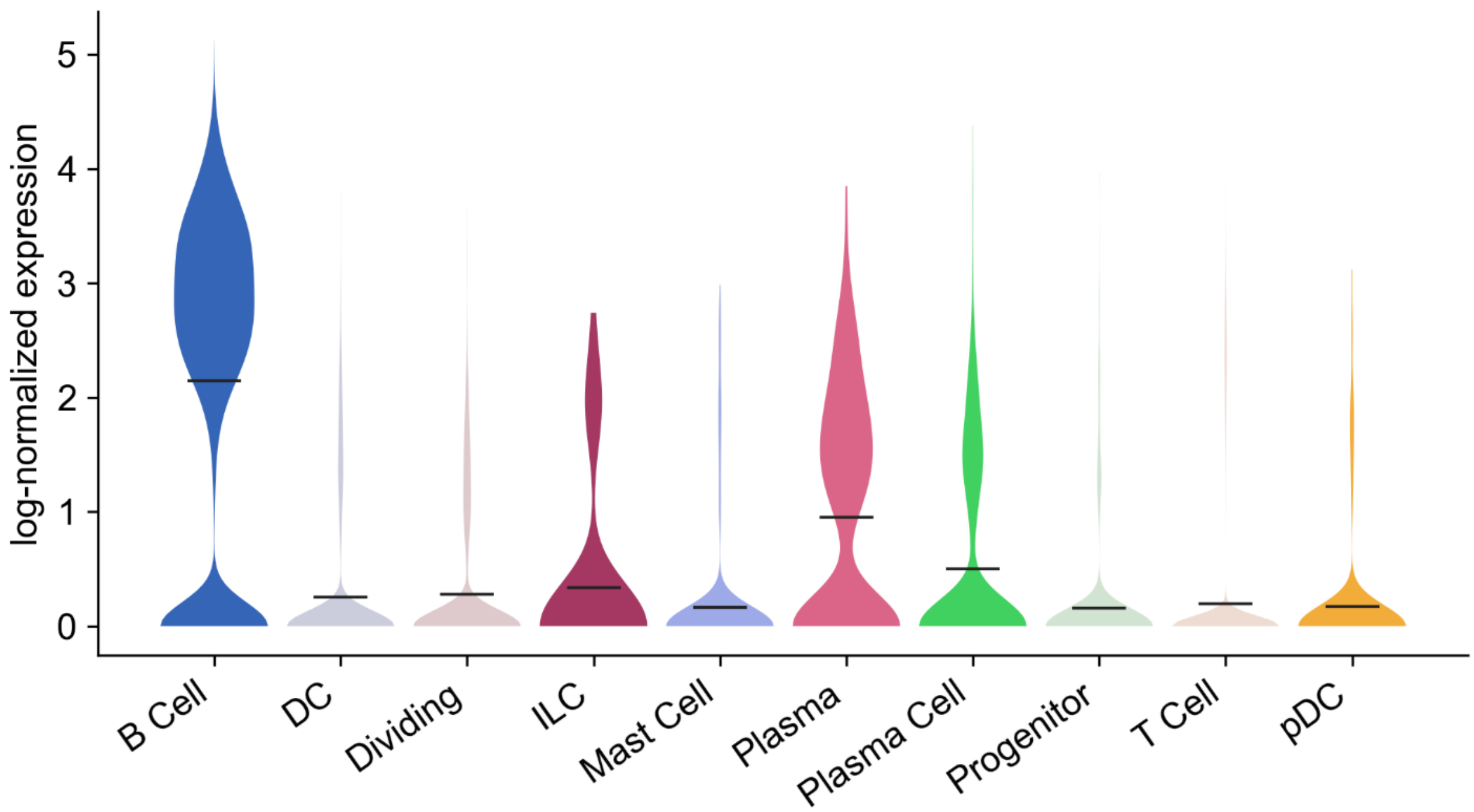

Figure S8: Violin plots of log-normalized expression on the green-monkey dataset.

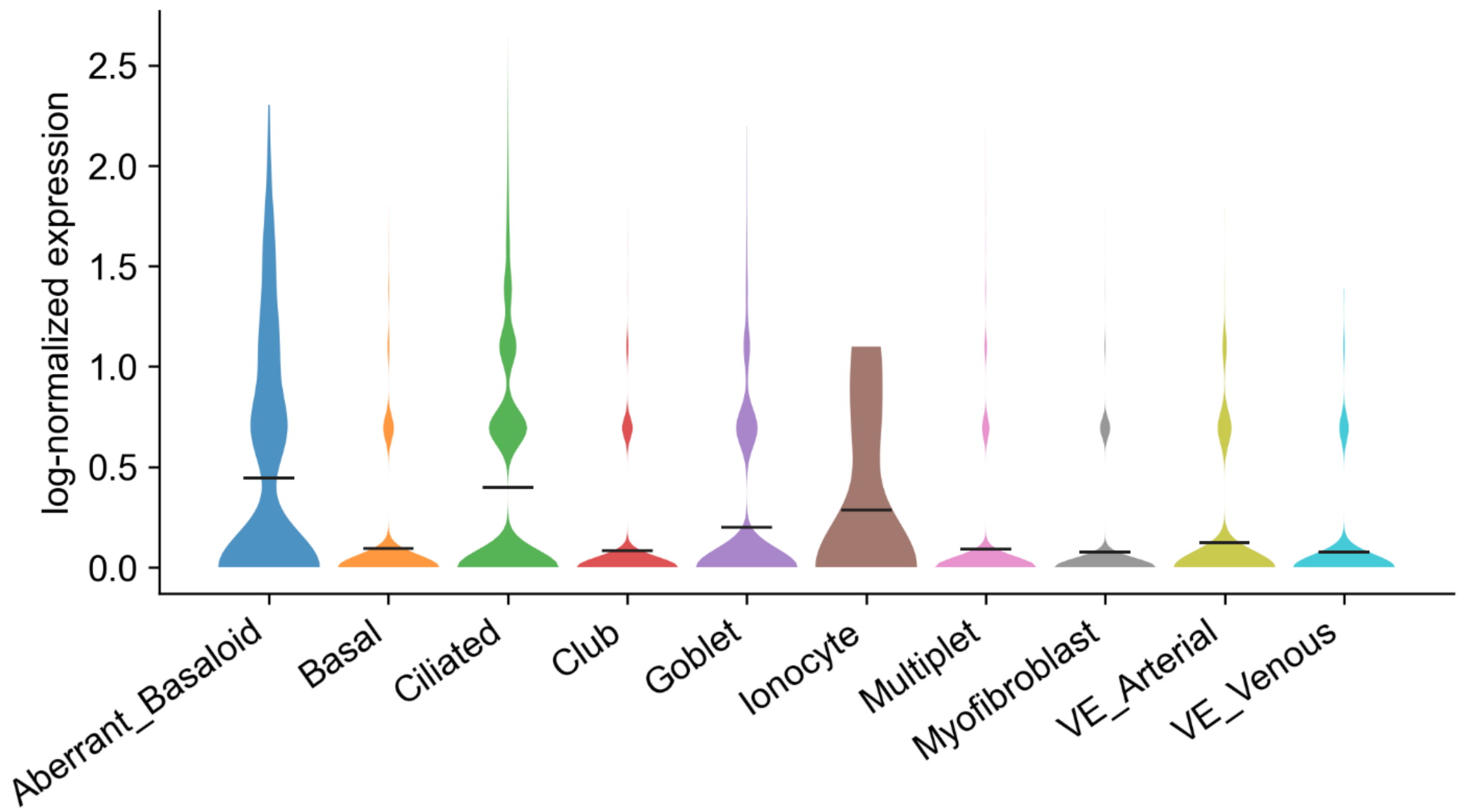

Figure S9: Violin plots of log-normalized expression on the human-lung dataset.

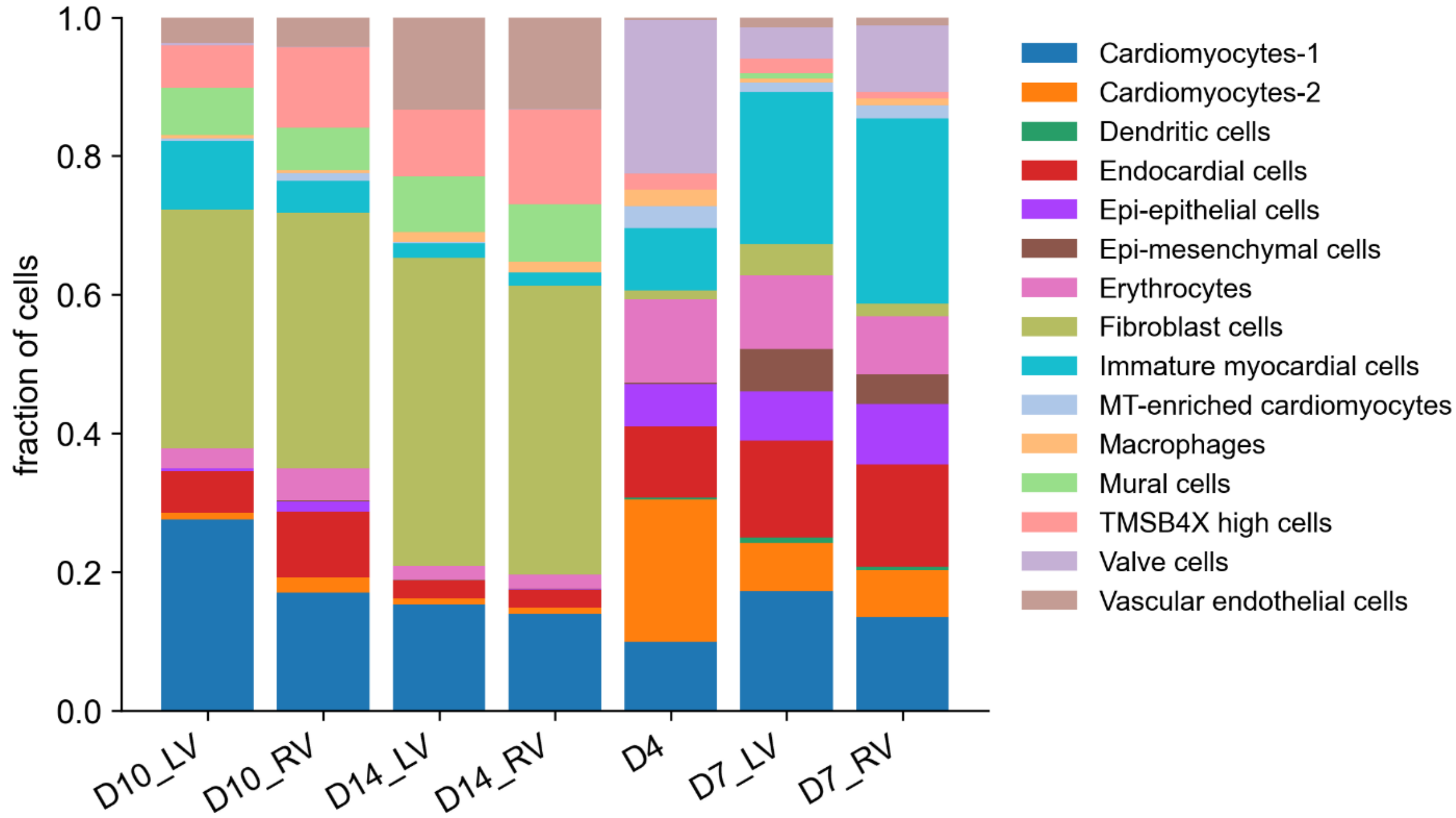

Figure S10: Stacked bar plot of cell-type fractions (14 annotated types) across seven chicken-heart samples (D4; D7, D10, D14 each split into LV/RV).

| Cardiomyocytes-1 | Cardiomyocytes-2 | Dendritic cells | Endocardial cells | Epi-epithelial cells | Epi-mesenchymal cells |
|---|---|---|---|---|---|
| *MYL2* | *HBA1* | *IFI6* | *FN1* | *MDK* | *POSTN* |
| *MYL3* | *HBZ* | *ITGB3* | *POSTN* | *TGFB2* | *TCF21* |
| *TPM4* | *ACTC1* | *CSTA* | *CDH5* | *TCF21* | *UGDH* |
| *TNNT2* | *TNNC1* | *RAC2* | *ND6* | *NREP* | *CDH2* |
| *ACTC1* | *MYL2* | *TMSB4X* | *ENG* | *COL3A1* | *MYH9* |
| *TNNC1* | *GAPDH* | *MPL* | *CST3* | *ALDH1A2* | *FN1* |
| *HSPB1* | *TPM4* | *LDHA* | *JCAD* | *FBLN1* | *DCN* |
| *TNNI1* | *MYH15* | *MYH9* | *RARRES1* | *MYH10* | *ENG* |

Figure S11: Marker/DE table snapshot for a selected cell-type group on the chicken-heart dataset.

| B Cell | Cilated | DC | Dividing | Fibroblasts | ILC |
|---|---|---|---|---|---|
| *CD79B* | *PIFO* | *VMO1* | *STMN1* | *TIMP1* | *NFKBIA* |
| *MS4A1* | *ALDH1A1* | *MNDA* | *TOP2A* | *SERPING1* | *IL7R* |
| *RPS8* | *AGR3* | *CDKN1A* | *HMGB2* | *MGP* | *ID2* |
| *IGHM* | *RSPH1* | *BCL2A1* | *MKI67* | *DCN* | *TNFAIP3* |
| *RPS3A* | *TSPAN1* | *HLA-DRA* | *CENPF* | *CALD1* | *ZFP36* |
| *RPS5* | *LKAAEAR1* | *CD86* | *SMC4* | *ADIRF* | *TYROBP* |
| *IGKC* | *CFAP53* | *LSP1* | *DUT* | *CST3* | *CD69* |
| *C12orf57* | *CCDC146* | *IDO1* | *SMC2* | *IGFBP5* | *GPR183* |

Figure S12: Marker/DE table snapshot for a selected cell-type group on green-monkey dataset.

| ATI | ATII | Aberrant_Basaloid | B | B_Plasma | Basal | Ciliated | Club |
| --- | --- | --- | --- | --- | --- | --- | --- |
| EMP2 | ANK3 | PTPRK | TNFRSF13C | ANKRD28 | EYA2 | LRRIQ1 | NEDD4L |
| LIMCH1 | ACOXL | PON2 | BANK1 | GNG7 | CACHD1 | RSPH1 | SFTPB |
| LMO7 | NEDD4L | PATJ | ARHGAP24 | POU2AF1 | FAF1 | CFAP299 | MECOM |
| SUSD2 | PDE4D | SMURF1 | SMAP2 | SMAP2 | ANK3 | MS4A8 | PATJ |
| CAV1 | ZNF385B | MYO6 | ADAM28 | SSR4 | NEDD4L | MAPK10 | ATP13A4 |
| AGER | SFTPB | MT-CYB | MGAT5 | FNDC3A | MCC | AGBL4 | GPRC5A |
| SLC39A8 | MACROD2 | PARD3 | ANKRD44 | RALGPS2 | XPR1 | SPATA17 | MCU |
| SCEL | ABCA3 | DST | CD79A | ELL2 | ACTN4 | CFAP54 | MGST1 |
| LAMA3 | AUTS2 | MYOF | BLK | HERPUD1 | SLC22A23 | DNAI2 | MT-ATP6 |
| CADM1 | ERBB4 | DOCK1 | AFF3 | USP48 | UGT1A8 | RFX3 | SHROOM3 |

| DC_Langerhans | DC_Mature | Fibroblast | Goblet | ILC_A | ILC_B | Ionocyte | Lymphatic |
| --- | --- | --- | --- | --- | --- | --- | --- |
| SLC8A1 | CSF2RA | DCN | BMPR1B | KLRD1 | KLRB1 | STAP1 | TFPI |
| CD74 | CERS6 | C1S | MGST1 | MCTP2 | IL7R | TFCP2L1 | TFF3 |
| HLA-DRA | LAMP3 | FBLN1 | CXCL17 | PARP8 | RUNX2 | PDE1C | PPFIBP1 |
| STK17B | FNBP1 | SERPING1 | KIAA1217 | FYN | TNFRSF18 | CD9 | GNG11 |
| HLA-DPB1 | CD83 | C1R | NEBL | SYTL3 | SYTL3 | ATP6V0B | CAVIN2 |
| LST1 | STK4 | LAMA2 | CAPN8 | TXK | CD69 | ROR1 | MMRN1 |
| HLA-DQA1 | SMAP2 | LRP1 | FHIT | KLRC4-KLRK1 | FXYD5 | GPC6 | AKAP12 |
| HLA-DPA1 | BIRC3 | SERPINF1 | NEDD4L | AOAH | CD7 | KCNMA1 | TBX1 |
| ELMO1 | PTPN1 | COL6A2 | EHF | KLRC1 | CTSW | CLCNKA | PKHD1L1 |
| HLA-DQB1 | GRK3 | CFD | MT-CO1 | CD96 | XCL2 | SCNN1B | STOX2 |

| Macrophage | Macrophage_Alveolar | Mast | Mesothelial | Multiplet | Myofibroblast | NK | PNEC |
| --- | --- | --- | --- | --- | --- | --- | --- |
| SLC11A1 | SLC11A1 | SLC24A3 | LSAMP | SNTB1 | LAMA2 | NKG7 | NRG1 |
| ZEB2 | OLR1 | KIT | C3 | SLC11A1 | COL6A3 | CD247 | SDK1 |
| LAPTM5 | C1QA | IL1RL1 | RBFOX1 | MT-ND5 | CDH11 | PRF1 | NKAIN2 |
| CEBPB | MS4A6E | CPA3 | HAS1 | ZEB2 | PRKG1 | KLRD1 | MTMR7 |
| TYROBP | LAPTM5 | TPSB2 | WT1 | LAPTM5 | LUM | AOAH | RIMS2 |
| OLR1 | GRN | TPSAB1 | ROR2 | DPYD | RORA | CST7 | MYT1L |
| FCER1G | PSAP | CHN2 | PDPN | SAMHD1 | RARRES2 | PIP4K2A | ATP8A2 |
| MS4A6E | MSR1 | HPGDS | C4B | MT-ND6 | FBLN1 | GZMB | NEBL |
| MGAT1 | GLUL | TNIK | CUX1 | MT-CO2 | LRP1 | GNLY | PTPRN2 |
| GLUL | SNTB1 | GATA2 | MTHFD1L | MT-CO1 | MMP2 | KLRF1 | PCDH9 |

| Pericyte | SMC | T | T_Cytotoxic | T_Regulatory | VE_Arterial | VE_Capillary_A | VE_Capillary_B |
| --- | --- | --- | --- | --- | --- | --- | --- |
| PTEN | MYL9 | ITK | CCL5 | CD247 | PTPRB | ADGRL2 | EPAS1 |
| NID1 | PPP1R14A | CNOT6L | KLRC4-KLRK1 | SKAP1 | LDB2 | HPGD | BTNL9 |
| PDGFRB | SELENOM | PBX4 | PPP2R5C | ICOS | EPAS1 | CLDN5 | LDB2 |
| COL4A2 | PRKG1 | PARP8 | PARP8 | PBX4 | EFNB2 | IFI27 | ADGRL2 |
| PRKG1 | TPM2 | ARHGAP15 | FYN | DUSP16 | PTPRM | EDNRB | FCN3 |
| COL4A1 | MYH11 | CRYBG1 | CD3D | BATF | ERG | AFF3 | TMEM100 |
| CALD1 | LMOD1 | BCL11B | CNOT6L | CD3D | TMTC1 | HLA-E | PTPRB |
| NOTCH3 | SOD3 | CD3D | PTPRC | PTPRC | PREX2 | EMCN | NOSTRIN |
| EBF1 | DSTN | CAMK4 | AOAH | BCL11B | CALCRL | GALNT18 | VWF |
| MYO1B | COX7A1 | IL7R | SYTL3 | CD28 | IL33 | STXBP6 | PECAM1 |

| VE_Peribronchial | VE_Venous | cDC1 | cDC2 | cMonocyte | ncMonocyte | pDC |
| --- | --- | --- | --- | --- | --- | --- |
| SPARCL1 | VWF | HLA-DPB1 | HLA-DRA | THBS1 | LST1 | FCHSD2 |
| SASH1 | PTPRB | HLA-DPA1 | HLA-DPB1 | SAMSN1 | FCER1G | IRF4 |
| TCF4 | ANO2 | CSF2RA | CD74 | SIPA1L1 | AIF1 | INPP4A |
| ADGRL4 | LIFR | HLA-DRA | HLA-DQA1 | FCN1 | PAG1 | PDE4B |
| PECAM1 | MRTFB | WDFY4 | TYROBP | SRGN | CTSS | SEC61B |
| ZNF385D | ACKR1 | HLA-DQA1 | CSF2RA | C5AR1 | SMAP2 | RUBCN |
| PITPNC1 | EPAS1 | CD74 | SIPA1L1 | CTSS | SAT1 | GZMB |
| ERG | PECAM1 | PFKFB3 | HLA-DRB1 | VCAN | SAMSN1 | SEL1L3 |
| VWF | PRSS23 | HLA-DRB1 | HLA-DPA1 | NAMPT | COTL1 | RHEX |
| MAGI1 | RAMP3 | CLEC7A | SRGN | SLC11A1 | TBXAS1 | ANKRD11 |

Figure S13: Marker/DE table snapshot for a selected cell-type group on the human-lung dataset.